\documentclass[aps,preprint,pre,eqsecnum]{revtex4-1} 

\usepackage{natbib}
\usepackage{graphicx}
\usepackage{subfigure}
\usepackage{amssymb}
\usepackage{amsfonts}
\usepackage{amsmath}
\usepackage{amsbsy}
\usepackage{mathrsfs} 
\usepackage{color}
\usepackage{gensymb} 
\usepackage{hyperref} 
\usepackage{soul}
\usepackage{lineno}
\usepackage{esint}
\usepackage{gensymb}
\providecommand\bu{\mathbf{u}}

\providecommand\bzero{\mathbf{0}}

\providecommand\be{\mathbf{\hat{e}}}

\newcommand{\pd}[2]{\frac{\partial #1}{\partial #2}}

\newcommand{\ub}[1]{^{({#1})}}

\begin{document}
\title{Boundary-layer effects on electromagnetic and acoustic extraordinary transmission through narrow slits}
\author{Rodolfo Brand\~ao*} \author{Jacob R. Holley*} \author{Ory Schnitzer}
\affiliation{Department of Mathematics, Imperial College London, SW7 2AZ, UK}
\begin{abstract}
We study the problem of resonant extraordinary transmission of electromagnetic and acoustic waves through subwavelength slits in an infinite plate, whose thickness is close to a half-multiple of the wavelength. We build on the matched-asymptotics analysis of Holley \& Schnitzer (Wave Motion, \textbf{91} 102381, 2019), who considered a single-slit configuration assuming an idealised formulation where dissipation is neglected and the electromagnetic and acoustic problems are analogous. We here extend that theory to include thin dissipative boundary layers associated with finite conductivity of the plate in the electromagnetic problem and viscous and thermal effects in the acoustic problem, considering both single-slit and slit-array configurations. 
By considering a distinguished boundary-layer scaling where dissipative and diffractive effects are comparable, we develop accurate analytical approximations that are generally valid near resonance; the electromagnetic-acoustic analogy is preserved up to a single physics-dependent parameter that is provided explicitly for both scenarios. The theory is shown to be in excellent agreement with GHz-microwave  and kHz-acoustic  experiments in the literature. 
\end{abstract}
\maketitle

\section{Introduction}\label{sec:into}
The phenomenon of extraordinary transmission of wave energy through small openings is key to the operation of numerous structured devices used for wave manipulation, ranging from traditional filters and gratings to modern metasurfaces and metamaterials \cite{Ebbesen:98,Porto:99}. A basic example is resonant transmission through narrow slits in an infinite plate (single slits or slit arrays), where the formation of standing waves in the slits gives rise to so-called Fabry--P\'erot transmission resonances \cite{Takakura:01,Yang:02,Garcia:03,Suckling:04,Lawrence:04,Bravo:04,Lindberg:04,Christensen:08, Ward:15,Ward:16,Moleron:16,Lin:17,Holley:19}. 

In the simplest version of this problem, there exists a precise electromagnetic-acoustic analogy in which the wave is represented by a two-dimensional scalar field governed by the Helmholtz equation along with a homogeneous Neumann condition on the plate boundary. There are no dissipative mechanisms in this idealised analogue model. In the electromagnetic scenario, it is valid for transverse-magnetic (TM) polarised waves, assuming that the plate is perfectly conducting. In the acoustic scenario, it is required that the plate is rigid and that viscous and thermal effects are negligible. This analogy naturally extends with  appropriate assumptions to other physical scenarios such as  water waves \cite{Evans:18}. 

For a single-slit configuration, the idealised analogue model predicts enhanced transmission at a set of resonance frequencies, which owing to diffractive effects are slightly lower than the  standing-wave frequencies of the slit calculated with end effects ignored \cite{Takakura:01,Holley:19}. For a slit-array configuration, this resonant enhancement gives rise to perfect transmission \cite{Takakura:01}. 

The assumptions underlying this idealised model are often unrealistic. Thus, electromagnetic experiments with metallic plates in the GHz-microwave regime, together with numerical simulations, have shown that owing to the  finite conductivity of the plate the resonant transmission peaks are diminished and shifted to lower frequencies (on top of the diffractive shifts predicted by the idealised model) \cite{Suckling:04}. Qualitatively similar effects occur in the acoustic problem, owing to thermal and viscous effects, as demonstrated by experiments and simulations for kHz frequencies in air \cite{Ward:15}. In both physical scenarios, these discrepancies with the analogue model are attributed to the effects of dissipative boundary layers that are thin compared to the subwavelength width of the slits. In the electromagnetic scenario, the boundary layer, aka skin, lies within the plate. In contrast, the thermoviscous boundary layer in the acoustic scenario lies in the exterior fluid domain.  

Our goal is to analytically investigate the above electromagnetic and acoustic boundary-layer effects, for both single-slit and slit-array configurations. In accordance with the relevant experiments, we shall take the incident field to be a plane wave propagating perpendicular to the plate. Furthermore, in the acoustic scenario we shall assume that the fluid is a viscous and thermally conducting ideal gas and that the plate is rigid and isothermal; in the electromagnetic scenario we shall assume the TM polarisation and that the plate is metallic. Our aim is to derive an asymptotic description based on the subwavelength smallness of the slit width in comparison with the wavelength-scale plate thickness, starting from first principles: the macroscopic Maxwell equations in the electromagnetic scenario and the linearised Navier--Stokes and energy equations in the acoustic scenario. In  appearance, these two formulations are very different. Our analysis will show, however, that the physical analogy between the idealised electromagnetic and acoustic problems can be essentially carried over to the corresponding dissipative problems.

We shall extensively build on the analysis of Holley and Schnitzer \cite{Holley:19}, who used the method of matched asymptotic expansions \cite{Hinch:91,Van:pert} to analyse the idealised problem in the case of a single slit. There are two important elements in that analysis which distinguish it from previous analyses of the idealised problem; these will be crucial here too. The first is that the analysis focuses on ``near-resonance'' regimes in frequency and parameter space. In the idealised problem, these regimes are defined by the smallness of the slit width in comparison to the plate thickness, together with the related proximity of the frequency to one of the standing-wave frequencies of the slit. In the present context, we also require that the boundary layers are thin compared to the slit width, so that the resonances are not strongly damped owing to dissipative effects. Specifically, without loss of generality we shall consider a distinguished limit where boundary-layer and diffractive effects are comparable in order of magnitude. The second important element is the systematic treatment of diffractive end effects via matched asymptotics and conformal mapping techniques. The asymptotic results derived in \cite{Holley:19} corrected previous approximations \cite{Takakura:01} and were shown to be in excellent agreement with  numerical solutions of the idealised problem and with a subset of the microwave experiments in \cite{Suckling:04} for which dissipative effects are minimal. The motivation for this work is partially to extend this agreement to the remaining experimental data in \cite{Suckling:04} and to the acoustic experiments in \cite{Ward:15}.

The rest of the paper is structured as following. In \S\ref{sec:form} we formulate the electromagnetic and acoustic problems for transmission of a normally incident plane wave through a single slit. In \S\ref{sec:single} we asymptotically analyse these problems, jointly for the most part. In \S\ref{sec:array} we extend the theory to a periodic array of slits. For comparisons with experiments, see \S\ref{ssec:exp1} and \S\ref{ssec:exp2}. We give concluding remarks in \S\ref{sec:conc}.

\section{Formulation for a single slit}\label{sec:form}
\subsection{Electromagnetic problem} 
\label{ssec:electromagnetic}
\begin{figure}[t!]
	\begin{center}		\includegraphics[scale=0.38]{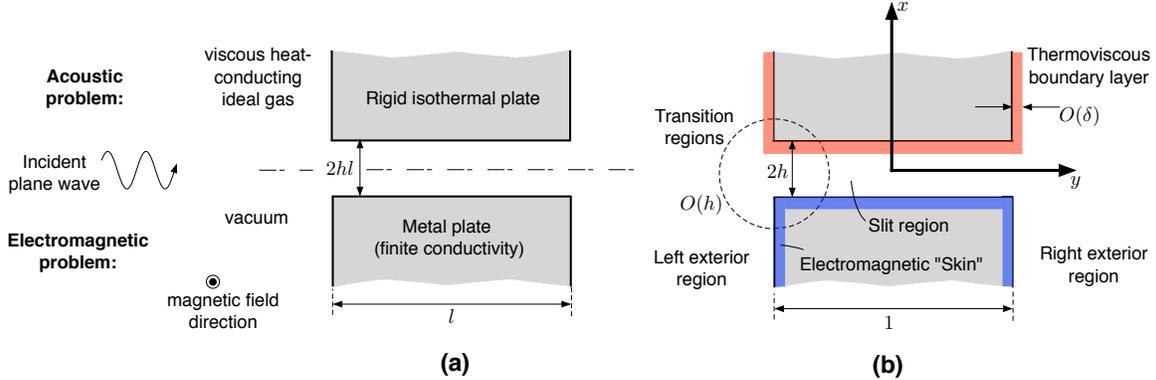}
		\caption{Transmission of a normally incident plane wave through a single slit in an infinite plate (upper half: acoustic problem, lower half: electromagnetic problem). (a) Dimensional schematic. (b) Dimensionless schematic showing the regions considered in the asymptotic analysis of \S\ref{sec:single}; the boundary-layer scaling $\delta=O(h^2)$ represents a distinguished limit where boundary-layer and diffraction effects are comparable.}
		\label{fig:schematic}
	\end{center}
\end{figure}

The electromagnetic problem is schematically depicted in Fig.~\ref{fig:schematic}(a). We consider an electromagnetic plane wave of  angular frequency $\omega$ normally incident on an infinite metallic plate of thickness $l$ and relative permittivity $\epsilon$. The plate is bisected by a single perpendicular slit of width $2hl$.  We consider a two-dimensional problem where the electric and magnetic fields are respectively parallel and perpendicular to the plane in Fig.~\ref{fig:schematic}, both fields being invariant in the perpendicular direction (TM polarisation). The background medium is assumed to be vacuum. 

In what follows we adopt a dimensionless convention where lengths are normalised by $l$. In particular, we will employ the dimensionless Cartesian coordinates $(x,y)$ shown in Fig.~\ref{fig:schematic}(b), along with the radial coordinates $r^{\pm}=\sqrt{x^2+(y\mp1/2)^2}$ measured from the centres of the slit apertures. We suppress the harmonic time variation 
$\exp(-i\omega t)$ in the usual way, 
such that fields are represented by complex-valued counterclockwise phasors. With this convention, $\epsilon$ is restricted to the second quadrant of the complex plane. We shall formulate the problem for the two-dimensional scalar fields $H$ and $\bar{H}$, the induction fields (normalised by the amplitude of the incident wave) in the background and plate domains, respectively. We also define the dimensionless frequency
\begin{equation}\label{Omega def}
\Omega=\frac{\omega l}{c},
\end{equation}
in which $c$ is the speed of light in vacuum. Note that $\Omega$ is of order unity in the case of interest where the wavelength is commensurate with the plate thickness.

The fields $H$ and $\bar{H}$ satisfy the reduced-wave equations
\refstepcounter{equation}
$$
\label{H eqs}
\nabla^2H+\Omega^2H=0, \quad \nabla^2\bar{H}+\epsilon\Omega^2\bar{H}=0,
\eqno{(\theequation{\mathit{a},\mathit{b}})}
$$
respectively in the background and plate domains. On the plate boundary we have the transmission conditions 
\refstepcounter{equation}
$$
\label{H bcs}
H=\bar{H}, \quad \frac{\partial H}{\partial n}=\frac{1}{\epsilon}\frac{\partial \bar{H}}{\partial n},
\eqno{(\theequation{\mathit{a},\mathit{b}})}
$$
where $\partial/\partial{n}$ represents the normal derivative. The problem is closed by specifying the incident field 
\begin{equation}\label{inc H}
H^{(i)}=e^{i\Omega y}
\end{equation}
along with the condition that the scattered field $H-H^{(i)}$ is outward radiating. 

An important dimensional length scale is the skin depth $l_s=\sqrt{2\rho/\omega\mu}$, 
where $\rho$ is the metal resistivity and $\mu$ the metal permeability, which is equal to the vacuum permeability. For a good conductor, $l_s$ is the characteristic scale on which  electromagnetic fields attenuate inside the metal. For our purposes, it is useful to define the dimensionless skin depth
\begin{equation}\label{delta s def}
\delta=\frac{l_s}{l}=\frac{1}{\Omega}\sqrt{\frac{2}{\epsilon''}},
\end{equation}
in which $\epsilon''$ denotes the imaginary component of $\epsilon$. The relative smallness of the skin depth is determined by the largeness of $\epsilon''$. In particular, for the GHz-microwave frequencies in the experiments in \cite{Suckling:04}, $\epsilon\approx i\epsilon''$ with $\epsilon''\simeq10^6-10^7$. 

\subsection{Acoustic problem}\label{ssec:acoustic}
We now formulate a sister acoustic problem, also depicted in Fig.~\ref{fig:schematic}(a). In this problem, the plate is rigid and isothermal (perfectly heat conducting), while the exterior domain is a viscous and heat-conducting ideal gas.  In contrast to the electromagnetic problem, the acoustic problem is confined to the exterior fluid domain. The governing equations are the continuity, momentum, energy and state equations linearised about an equilibrium state of density $\rho_0$, viscosity $\eta$, specific heat capacity at constant pressure $c_p$ and heat conductivity $\kappa$ \cite{Pierce:90}. We adopt the same dimensionless and phasor-field conventions as in the electromagnetic problem, with $c$ in \eqref{Omega def} now denoting the equilibrium speed of sound. Below, we formulate the acoustic problem for the dimensionless pressure $p$, temperature perturbation $T$ and velocity field $\mathbf{u}$ (respectively normalised by $p_{\infty}$, $p_\infty/(\rho_0c_p)$ and $p_\infty/(\omega\rho_0l)$). The reference pressure $p_{\infty}$ is associated with the incident plane wave, as  discussed below. 

The governing equations are the continuity equation
\begin{equation}\label{cont}
-i\Omega^2 p -i(\gamma - 1)\Omega^2(p-T)+\nabla\cdot\mathbf{u}=0,
\end{equation}
the momentum equation
\begin{equation}\label{mom}
-i\mathbf{u}=-\nabla p+\delta^2 \left[ \nabla^2\mathbf{u}+\frac{1}{3}\nabla\left( \nabla\cdot\mathbf{u} \right) \right]
\end{equation}
and the energy equation
\begin{equation}\label{energy}
i\left( p - T\right)=\text{Pr}^{-1}\delta^2\nabla^2T.
\end{equation}
Analogously to \eqref{delta s def}, we define the dimensionless viscous length scale
\begin{equation}\label{delta a def}
\delta=\frac{l_v}{l},
\end{equation} 
along with the Prandtl number $\mathrm{Pr}
=(l_v/l_t)^2$, in which $l_v=\sqrt{{\eta}/\rho_0\omega}$ and $l_t=\sqrt{{\kappa}/\rho_0\omega c_p}$ are dimensional viscous and thermal length scales, respectively.
The adiabatic index is denoted  $\gamma$. For air $\gamma\approx 1.4$ and $\text{Pr}\approx 0.71$. 

The above equations are supplemented by boundary and far-field conditions. On the plate boundary, the linearised flow field satisfies the no-slip condition
\begin{equation}\label{no slip}
\bu=\bzero,
\end{equation}
while the temperature perturbation satisfies the isothermal boundary condition
\begin{equation}\label{no temp}
T=0.
\end{equation} 
As in the electromagnetic problem, the acoustic problem is closed  by specifying a normally incident pressure plane wave and requiring the corresponding scattered fields to be outward propagating. Strictly speaking, plane-wave solutions of \eqref{cont}--\eqref{energy} attenuate exponentially; thus, we define the reference value $p_{\infty}$ as the magnitude of the plane wave at $y=0$. For small $\delta$, however, sound attenuation in the bulk of the fluid occurs on a length scale large compared to the plate thickness \cite{Pierce:90}. For this reason, it will suffice that at a fixed position the incident pressure field $p\ub{i}$ satisfies
\begin{equation}\label{inc p limit}
p^{(i)}\to e^{i\Omega y} \quad \text{as} \quad  \delta\to0.
\end{equation}
This limit is identical to the exact incident field \eqref{inc H} in the electromagnetic problem.

\subsection{Near-resonance limit} \label{ssec:limit}
Following the analysis of the idealised problem in \cite{Holley:19}, we shall consider the near-resonance regime where $h\ll1$ and $\Omega-\bar{\Omega}=O(h)$, with
\begin{equation}
\bar{\Omega}=m\pi, \quad m=1,2,3,\ldots \label{FP}
\end{equation} 
the unperturbed standing-wave, or Fabry--P\'erot, frequencies of the slit (i.e., ignoring end-correction and dissipative effects). For this purpose, it is convenient to write 
\begin{equation}\label{Omegap def}
\frac{\Omega-\bar{\Omega}}{\bar{\Omega}}\equiv 2h\Omega'
\end{equation}
with $\Omega'$ fixed as $h\to0$. For this regime, it was shown that the wave field within the slit is approximated by the corresponding one-dimensional standing wave, except close to the slit ends, of magnitude $O(1/h)$ --- singularly large compared to the $O(1)$ magnitude of the slit field off-resonance and the $O(1)$ magnitude of the incident wave; concomitantly, the field diffracted from the slit ends was shown to be enhanced from $O(h)$ to $O(1)$. 

It is clear that dissipative effects can only widen the resonances whilst diminishing the above magnitude scalings. In what follows, we shall specifically consider the distinguished boundary-layer scaling $\delta=O(h^2)$ where, on the one hand, dissipative effects are important at leading order, while, on the other hand, the near-resonance scalings are the same as in the idealised problem. This distinguished scaling, which we shall see holds for both the electromagnetic and acoustic problems, represents a balance between diffractive energy leakage from the slit and dissipation. 

In light of the above, we rescale the dimensionless boundary-layer thickness as 
\begin{equation}\label{delta p def}
\delta= h^2\delta',
\end{equation}
with $\delta'$ fixed as $h\to0$. 
In the electromagnetic problem, $\delta$ is implicit in the problem formulation. Thus, in that problem it will be more convenient to hold fixed the complex-valued parameter 
\begin{equation}\label{alpha def}
\alpha=h^2\epsilon^{1/2},
\end{equation}
where the principal branch of the square-root function is assumed.

\section{Asymptotic analysis for a single slit}\label{sec:single}
\subsection{Matched asymptotic expansions}
In this section we study the single-slit problems formulated in \S\ref{sec:form}, considering  the near-resonance limit process described in \S\ref{ssec:limit}. Following \cite{Holley:19}, we shall employ the method of matched asymptotic expansions whereby the physical domain is conceptually decomposed into  overlapping regions interconnected by asymptotic matching rules \cite{Hinch:91,Van:pert}. The regions employed in our analysis are schematically shown in Fig.~\ref{fig:schematic}(b). These include the ``bulk'' left and right exterior regions, a slit region and two transition regions close to the slit ends, as well as boundary-layer regions  respectively inside and outside the plate in the electromagnetic and acoustic problems. The term bulk is used to indicate characteristic dimensions that are large compared to the boundary-layer thickness. Matching in-between the bulk regions will be carried out based on results from \cite{Holley:19}. In the acoustic problem, further matching will be needed in order to connect the overlapping boundary-layer and bulk regions. In contrast, in the electromagnetic problem the boundary layers are more simply coupled to the bulk regions, through boundary conditions \eqref{H bcs}. 

\subsection{Boundary-layer effects: scalings} \label{ssec:scalings}
It is constructive to preface the asymptotic analysis by deriving scaling rules describing the effects of the thin boundary layers on wave propagation in the bulk exterior, slit and transition regions. These scaling rules will help to streamline the analysis, motivate a joint treatment of the electromagnetic and acoustic problems and rationalise the distinguished boundary-layer scaling \eqref{delta p def}.

The scaling argument is particularly simple in the electromagnetic scenario, in which the electromagnetic boundary layer lies inside the metal plate. Let the scaling of $H$ in a given bulk region be $\mathscr{H}$; then the field $\bar{H}$ in the electromagnetic boundary layer is comparable in magnitude, attenuating into the metal domain on the boundary-layer scale $\delta$. With this construction, the transmission boundary condition (\ref{H bcs}b) together with \eqref{delta s def} implies a bulk perturbation of order 
\begin{equation}
\mathscr{H}'=\delta  \Delta_{\perp}\mathscr{H},
\end{equation}
where $\Delta_{\perp}$ denotes the characteristic length scale of the bulk field normal to the boundary (unity for the exterior regions and $h$ for the slit and transition regions). We assume that, as in the near-resonance lossless analysis \cite{Holley:19}, $\mathscr{H}=h^{-1}$ in the slit region and unity in the exterior and transition regions. Then, with the distinguished scaling $\delta=O(h^2)$, we find $\mathscr{H}'=h^2,h^2$ and $h^3$ for the exterior, slit and transition regions, respectively.

Consider next the acoustic problem, in which there are viscous and thermal boundary layers on the external side of the plate boundary. The tangential component of the velocity field varies across the boundary layer, from the bulk value, of order $\mathscr{U}$, say, to zero at the boundary; the pressure, of order $\mathscr{P}$, say, is approximately uniform across the layer. Furthermore, the temperature varies from its bulk value, on the order of $\mathscr{P}$, to zero on the boundary. With this construction, the continuity equation \eqref{cont} implies bulk perturbations of the \textit{normal} velocity component of order $\mathscr{U}_V'=(\delta/\Delta_{\parallel})\mathscr{U}$ and $\mathscr{U}_T'=\delta\mathscr{P}$ owing to viscous and thermal effects, respectively, wherein $\Delta_{\parallel}$ is the characteristic scale of the bulk field parallel to the boundary (unity for the exterior and slit regions and $h$ for the transition regions). Since in the bulk velocity scales like the pressure gradient we have $\mathscr{U}=\mathscr{P}/\Delta_{\parallel}$ and $\mathscr{U}_{V/T}=\mathscr{P}_{V/T}'/\Delta_{\perp}$, where $\mathscr{P}_{V}$ and $\mathscr{P}_T$ are the orders of the bulk pressure perturbations owing to viscous and thermal effects, respectively. We therefore find the estimates 
\begin{equation}
\mathscr{P}_V'=\frac{\delta {\Delta}_{\perp}} {{\Delta_{\parallel}}^2}\mathscr{P} \quad \text{and} \quad \mathscr{P}_T'=\delta\Delta_{\perp}\mathscr{P}.
\end{equation}
With $\mathscr{P}=h^{-1}$ in the slit region and unity in the exterior and transition regions, we find $\mathscr{P}_V'=h^2,h^2$ and $h$ in the exterior, slit and transition regions, respectively; for $\mathscr{P}_T'$ we find   the same orders as for the induction field in the electromagnetic problem.

The relative smallness of these bulk perturbations may appear to contradict our  claim that the assumed scaling $\delta=O(h^2)$ represents a distinguished limit where boundary-layer effects are leading order. While these perturbations are indeed negligible in the exterior and transition bulk regions, we know from the analysis of the idealised problem in \cite{Holley:19} that $O(h^2)$ perturbations in the slit region affect the leading $O(h^{-1})$ field in the slit. It is this observation, together with the above scaling rules, by which we have identified the distinguished scaling \eqref{delta p def}. We note that an alternative scaling analysis could be carried out based on energy-dissipation arguments \cite{Brandao:20H}. 

\subsection{Exterior regions}\label{ssec:exterior}
Armed with these scaling results, we begin our analysis with the exterior regions, which are defined by the limit process discussed in \S\ref{ssec:limit} together with the additional specification that the coordinates $(x,y)$ are held fixed. We shall distinguish between a left exterior region corresponding to the half-space $y<-1/2$ and a right exterior region corresponding to the half-space $y>1/2$. For the sake of considering the electromagnetic and acoustic problems simultaneously, we pose the asymptotic expansions
\begin{equation}
H,p \sim \varphi^{\pm}(x,y) \quad \text{as} \quad h\to0,
\end{equation}
the plus-minus signs indicating the right and left exterior regions, respectively. 

The leading-order fields $\varphi^\pm$ satisfy the Helmholtz equations \begin{equation}\label{helm exteriors}
\nabla^2\varphi^{\pm}+\bar{\Omega}^2\varphi^{\pm}=0
\end{equation}
in the respective half-spaces. In the electromagnetic problem, \eqref{helm exteriors} readily follows from a leading-order balance of (\ref{H eqs}a). In the acoustic problem, it follows from combining the leading-order balances of \eqref{cont}--\eqref{energy}. Note that the dimensionless frequency is approximated at leading order by $\bar{\Omega}$ in accordance with the frequency rescaling \eqref{Omegap def}. 

The Helmholtz equations \eqref{helm exteriors} are supplemented by effective boundary conditions.  
At $y=\pm1/2$,  the scaling results of \S\ref{ssec:scalings} imply that in the exterior regions boundary-layer effects are negligible at $O(1)$. This implies the Neumann boundary conditions 
\refstepcounter{equation}
$$
\label{exterior Neumann}
\left.\pd{\varphi^-}{y}\right|_{y=-1/2}=0, \quad \left.\pd{\varphi^+}{y}\right|_{y=1/2}=0,\eqno{(\theequation{\mathit{a},\mathit{b}})}
$$
for $x\ne0$. The behaviour of $\varphi^{\pm}$ as $(x,y)\to(0,\pm{1/2})$ will be determined by asymptotic matching. As $|y|\to\infty$, the leading-order scattered fields $\varphi^{\pm}-\varphi\ub{i}$ satisfy an outward-radiation condition with respect to the incident field [cf.~\eqref{inc H} and \eqref{inc p limit}]
\begin{equation}\label{incident exterior}
\varphi\ub{i}=e^{i\bar{\Omega}y}.
\end{equation}

Consider the left exterior region. The solution for $\varphi^-$ is obtained by superposing the incident plane wave \eqref{incident exterior}, a reflected plane wave, whose phase is adjusted to satisfy  condition (\ref{exterior Neumann}a), and the fundamental singular solutions of the Helmholtz equation \eqref{helm exteriors} with origin at $(x,y)=(0,-1/2)$. This singular solution, which represents an outward-radiating cylindrical wave, is proportional to $\mathcal{H}_0(\bar{\Omega}r^-)$, wherein $\mathcal{H}_0$ is the zeroth-order Hankel function \cite{Abramowitz:book}; it is clearly compatible with (\ref{exterior Neumann}a). An asymptotic property of this function that will be important later is
\begin{equation}\label{hankel sing}
\mathcal{H}_0(s)\sim \frac{2i}{\pi}\ln s+\frac{2i}{\pi}\left(\gamma_E-\ln 2\right)+1 + o(1) \quad \text{as} \quad s\to0,
\end{equation}
where $\gamma_E=0.5772\ldots$ is the Euler--Mascheroni constant. Higher-order singular solutions, formed by derivates of this fundamental solution, can be eliminated via matching \cite{Holley:19}. We accordingly find
\begin{equation}\label{exterior left}
\varphi^-= e^{i\bar{\Omega}y}+Re^{-i\bar{\Omega}y}+Q^-\mathcal{H}_0(\bar{\Omega} r^-),
\end{equation}
where $R=e^{-i\bar{\Omega}}$ is a reflection coefficient and $Q^-$ is a diffraction coefficient to be determined.  

Similar considerations applied to the right exterior region give
\begin{equation}\label{exterior right}
\varphi^+= Q^+\mathcal{H}_0(\bar{\Omega} r^+),
\end{equation}
where $Q^+$ is a second diffraction coefficient to be determined. 

\subsection{Slit region}\label{ssec:slit}
Consider now the slit region, where the stretched transverse coordinate $X=x/h$ is held fixed instead of $x$. With that rescaling, we find the Helmholtz equation
\begin{equation}\label{slit helm}
\frac{1}{h^2}\pd{^2\Phi}{X^2}+\pd{^2\Phi}{y^2}+{\bar\Omega}^2\left(1+4h\Omega'+4h^2{\Omega'}^2\right)\Phi\approx0
\end{equation}
for $|y|<1/2$ and $|X|<1$, where $\Phi(X,y)$ stands for $H$ in the electromagnetic problem and $p$ in the acoustic problem. In the former, \eqref{slit helm} follows from (\ref{H eqs}a) and \eqref{Omegap def} without approximation. In the latter, the governing equations \eqref{cont}--\eqref{energy} together imply \eqref{slit helm} with $O(\delta^2\Phi)$ error terms which are too small to affect the following analysis. 
 
The boundary conditions at $y=\pm1/2$ are to be determined by asymptotic matching with the transition and exterior regions. As for the boundary conditions at $X=\pm1$, the scalings derived in \S\ref{ssec:scalings} imply that the boundary layers at the top and bottom of the slit affect the bulk slit field only at high orders. Specifically,
\begin{equation}\label{slit BC scaling}
\pd{\Phi}{X}=O\left(h\delta\Phi\right) \quad \text{at} \quad X=\pm 1.
\end{equation}
The precise leading-order form of the right-hand side will be derived in \S\ref{ssec:skin} for the electromagnetic problem, by analysing the skin layers, and in \S\ref{ssec:thermoviscous} for the acoustic problem, by analysing the thermoviscous boundary layers. In the present subsection, we will simply quote these results (at a later point, as they depend on properties of $\Phi$ yet to be derived).

As already mentioned, we anticipate an $O(1/h)$ enhancement of the wave field in the slit. We accordingly pose the expansion
\begin{equation}\label{slit expansion}
\Phi= h^{-1}\Phi_{-1}(X,y)+\Phi_0(X,y) + h\Phi_1(X,y) +h^2\Phi_2(X,y) + \cdots \quad \text{as} \quad h\to0.
\end{equation}
The $O(1/h^3)$ and $O(1/h^2)$ balances of \eqref{slit helm} give the trivial relations
\begin{equation}
\pd{^2\Phi_{-1}}{X^2}=0,\quad \pd{^2\Phi_{0}}{X^2}=0.
\end{equation}
Since \eqref{slit BC scaling} implies the homogeneous Neumann conditions
\refstepcounter{equation}
$$
\label{neumann slit orders ab}
\pd{\Phi_{-1}}{X}=0, \quad \pd{\Phi_{0}}{X}=0  \quad \text{at} \quad X=\pm1, 
\eqno{(\theequation{\mathit{a},\mathit{b}})}
$$
we find that $\Phi_{-1}$ and $\Phi_0$ are independent of $X$, namely $\Phi_{-1}=\Phi_{-1}(y)$ and $\Phi_0=\Phi_0(y)$. Next, the $O(1/h)$ balance of \eqref{slit helm} gives
\begin{equation}\label{pre leading slit helm}
\pd{^2\Phi_1}{X^2}+\frac{d^2\Phi_{-1}}{dy^2}+{\bar\Omega}^2\Phi_{-1}=0.
\end{equation}
Integrating with respect to $X$ and using the homogenous Neumann condition
\begin{equation}\label{neumann slit orders c}
\pd{\Phi_1}{X}=0 \quad \text{at} \quad X=\pm1,
\end{equation} 
we find the one-dimensional Helmholtz equation
\begin{equation}\label{leading slit helm}
\frac{d^2\Phi_{-1}}{dy^2}+\bar{\Omega}^2\Phi_{-1}=0.
\end{equation}
It follows from \eqref{pre leading slit helm}--\eqref{leading slit helm}, in turn, that $\Phi_1=\Phi_1(y)$.

The Helmholtz equation \eqref{leading slit helm} is supplemented by the matching conditions
\begin{equation}\label{leading slit bcs}
\Phi_{-1}=0 \quad \text{at} \quad y=\pm1/2, 
\end{equation} 
which evidently follow from the relative smallness of the wave field in the exterior regions together with the regularity of solutions to \eqref{leading slit helm} \cite{Holley:19}.  
The homogeneous problem consisting of \eqref{leading slit helm} and \eqref{leading slit bcs} has nontrivial solutions only for $\bar\Omega=m\pi$, in which  the integer $m=1,2,\ldots$ corresponds to the order of the standing wave excited in the slit. We write these standing-wave solutions in the form 
\begin{gather}\label{homo solutions res loss}
\Phi_{-1}=\mathcal{A} \times \left\{\begin{array}{c}\cos(\bar{\Omega} y ) \\ \sin(\bar{\Omega} y )\end{array}\right\},
\end{gather}
where $\mathcal{A}$ is a complex-valued prefactor; we also introduce a notation to be used throughout the paper, where the upper element of the array corresponds to odd $m$ (even standing wave) and the lower to even $m$ (odd standing wave).

At this stage we need the right-hand side of \eqref{slit BC scaling}. With the knoweledge that $\Phi_{-1},\Phi_0$ and $\Phi_1$ are all independent of $X$, the boundary-layer analyses in \S\ref{ssec:skin} and \S\ref{ssec:thermoviscous} furnish the impedance-like conditions
\begin{equation}\label{blc}
\pd{\Phi_2}{X}=\pm\mathcal{C}\Phi_{-1} \quad \text{at} \quad X=\pm1,
\end{equation}
where $\mathcal{C}$ is a complex-valued parameter which is given in the electromagnetic scenario by \begin{equation}\label{C electromagnetic}
\mathcal{C}=\frac{i\bar{\Omega}}{h^2\epsilon^{1/2}}
\end{equation}
and in the acoustic scenario by 
\begin{equation}\label{C acoustic}
\mathcal{C}=\frac{\delta{\bar\Omega}^2}{h^2}\frac{1+i}{\sqrt{2}}\left(1+\frac{\gamma-1}{\sqrt{\text{Pr}}}\right),
\end{equation}
wherein $\delta$ is defined in \eqref{delta a def}. In \eqref{C electromagnetic} and \eqref{C acoustic}, $\epsilon$ and $\delta$, respectively, can be treated as constant over the near-resonance frequency interval.

Consider next the $O(1)$ balance of \eqref{slit helm},
\begin{equation}
\pd{^2\Phi_2}{X^2}+\frac{d^2\Phi_0}{dy^2}+{\bar\Omega}^2\Phi_0+4\Omega'{\bar\Omega}^2\Phi_{-1}=0.
\end{equation}
Integrating across the slit and using the boundary condition \eqref{blc} yields
\begin{equation}\label{slit helm corr}
\frac{d^2\Phi_{0}}{dy^2}+\bar{\Omega}^2\Phi_{0}=-\left(4\Omega'\bar{\Omega}^2+\mathcal{C}\right)\Phi_{-1}.
\end{equation}
We shall not need the detailed solution of \eqref{slit helm corr}. Rather, we derive relations between the leading-order amplitude $\mathcal{A}$ and the end values of $\Phi_0$, which will be required for asymptotic matching. To this end, we subtract the product of $\Phi_{0}$ and the complex conjugate of  \eqref{leading slit helm} from the product of  \eqref{slit helm corr} and the complex conjugate of $\Phi_{-1}$; integrating this difference between the two ends of the slit, we find
\begin{equation}
\left[\Phi_{0}\frac{d\Phi_{-1}^*}{dy}\right]_{-1/2}^{1/2}=\left(4\bar{\Omega}^2\Omega'+\mathcal{C}\right)\int_{-1/2}^{1/2}|\Phi_{-1}|^2dy.
\end{equation}
Substituting \eqref{homo solutions res loss} and using \eqref{FP} gives 
\begin{equation}\label{A loss}
\mathcal{A}= \frac{2\bar{\Omega}i^m}{4\Omega'\bar{\Omega}^2+\mathcal{C}}\times \left\{\begin{array}{c}i\Phi_{0}(1/2)+i\Phi_{0}(-1/2) \\ \Phi_{0}(1/2)-\Phi_{0}(-1/2) \end{array}\right\}.
\end{equation}

\subsection{Matching and transition regions}\label{ssec:aperture}
It remains to relate the approximations found in the exterior and slit regions. Given the logarithmic singularity of the leading-order exterior fields, these regions cannot be matched directly. Rather, it is necessary to consider intermediate ``transition regions'' at $O(h)$ distances from the slit ends. It is clear from the preceding analyses of the slit and exterior regions that $H$ and $p$ are $O(1)$ in these  regions. The scaling arguments in \S\ref{ssec:scalings} show that boundary-layer effects are not important at this order. The leading-order aperture problem is therefore the same as the one formulated and solved in \cite{Holley:19}, as are the details of the asymptotic matching between the transition regions and the slit and exterior regions. We shall simply quote the results of that analysis that are relevant here.

\begin{figure}[b!]
\begin{center}
\includegraphics[scale=0.4]{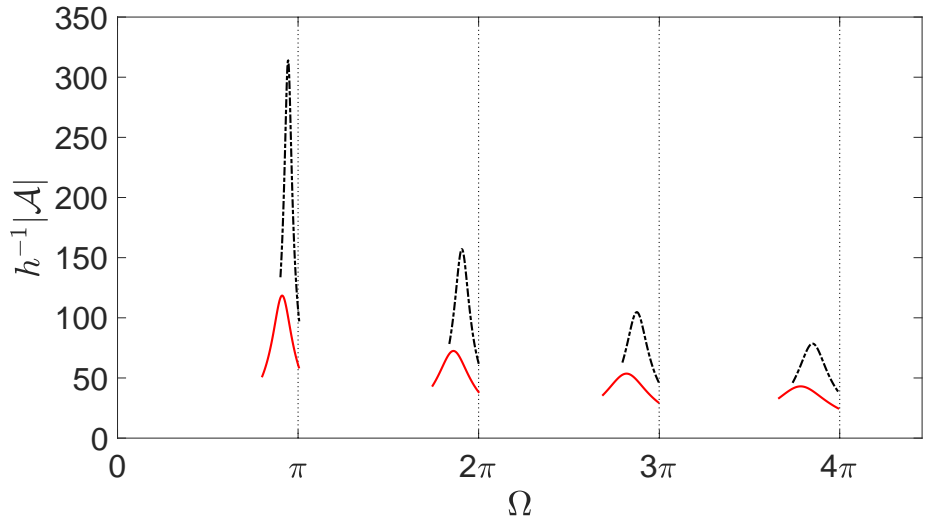}
\caption{Enhancement of the slit field predicted by the near-resonance approximation \eqref{A single} as a function of the dimensionless frequency $\Omega$. The slit aspect ratio is $h=0.01$. Black dashed lines: idealised problem without dissipation. Red solid lines: thermoviscous-acoustic scenario assuming air at $20\,\degree \mathrm{C}$ and slit length $l=35\mathrm{mm}$ \cite{Ward:15}. The vertical dotted lines mark the unperturbed standing-wave frequencies.}
\label{fig:freqresponse_single}
\end{center}
\end{figure}

The first result is a relation between the diffraction amplitudes $Q^{\pm}$ appearing in the $O(1)$ exterior fields and the rescaled amplitude $\mathcal{A}$ of the enhanced $O(h^{-1})$ standing wave in the slit. It can be written as
\begin{equation}\label{Q in A}
Q^{\pm} = i^m\bar{\Omega}\mathcal{A}\times \left\{\begin{array}{c} 1 \\ \mp i \end{array}\right\}. 
\end{equation}
The second result connects the diffraction coefficients $Q^{\pm}$ with the $O(1)$ discontinuity in the wave field across the aperture, with the singularity of the exterior field subtracted. This relation can be written as
\begin{equation}\label{relation 2}
\lim_{r^{\pm}\to0}\left(\varphi^{\pm}-\frac{2i}{\pi}Q^{\pm}\ln r^{\pm}\right)-\left.\Phi_0\right|_{y=\pm1/2}=iQ^{\pm}\left(\frac{2}{\pi}\ln \frac{1}{h}-\beta\right),
\end{equation}
where the parameter $\beta$ depends only on the geometry of the slit opening; for the right-angled aperture considered here, the conformal-mapping analysis in \cite{Holley:19} yields the value
\begin{equation}\label{beta def}
\beta = \frac{2}{\pi}\left(\ln\frac{4}{\pi}-1\right)\approx -0.4828.
\end{equation} 
Substituting the external fields $\varphi^{\pm}$ from \eqref{exterior left} and \eqref{exterior right}, and using the asymptotic relation \eqref{hankel sing}, we can solve \eqref{relation 2} for the end values of the $O(1)$ slit field:
\begin{equation}\label{Phi0 at ends}
\left.\Phi_0\right|_{y=\pm1/2}=Q^{\pm}\frac{2i}{\pi}\left(\ln\frac{2\bar{\Omega} h}{\pi}+\gamma_E-1-\frac{i\pi}{2}\right)+(1\mp 1)e^{-i\bar{\Omega}/2}.
\end{equation}

Note the logarithmic dependence upon $h$ in \eqref{Phi0 at ends}. This relation assumes an asymptotic convention where logarithmic orders are considered together  with the nearest algebraic order \cite{Hinch:91,Van:pert}. All small-$h$ expansions in this paper are to be interpreted in this manner.

\subsection{Frequency response}\label{ssec:resultsingle}
Solving \eqref{A loss}, \eqref{Q in A} and \eqref{Phi0 at ends} for the amplitude $\mathcal{A}$ yields 
\begin{equation}\label{A single}
\mathcal{A}= \frac{1/\bar{\Omega}}{\Omega'+\frac{\mathcal{C}}{4\bar{\Omega}^2}-\frac{2}{\pi}\left(\ln\frac{2\bar{\Omega} h}{\pi}+\gamma_E-1\right)+i}\times\left\{\begin{array}{c}i \\ -1 \end{array}\right\},
\end{equation}
which provides a closed-form approximation for the near-resonance frequency response. The corresponding results for the diffraction coefficients $Q^{\pm}$ can be obtained by substituting \eqref{A single} into \eqref{Q in A}. Recall that  \eqref{A single} was derived assuming the distinguished limit $\delta=O(h^2)$. For $\delta\ll h^2$, dissipation effects are negligible and we recover the results in \cite{Holley:19}. How about $\delta\gg h^2$? Since $\mathcal{C}=O(\delta/h^2)$, \eqref{A single} together with definition \eqref{Omegap def} for $\Omega'$ implies a sharp resonance as long as $\delta\ll h$ (boundary layer thin relative to slit). It can be verified that our results remain valid in this dissipation-dominated resonant regime; the resonance width becomes $\Omega-\bar{\Omega}=O(\delta/h)$ and the amplitude of the wave in the slit becomes $O(h/\delta)$. 

In Fig.~\ref{fig:freqresponse_single} we plot the enhancement factor $h^{-1}|\mathcal{A}|$ as a function of the dimensionless frequency $\Omega$. As an example, we set $h=0.01$ and compare the idealised case with the thermoviscous-acoustic scenario assuming the physical parameters of the experiments in \cite{Ward:15}. Note that the resonance peaks are shifted from the standing-wave values $\bar{\Omega}$, more so when dissipation is included. 

\subsection{Comparison with single-slit electromagnetic and acoustic experiments}\label{ssec:exp1}
\begin{figure}[t!]
	\begin{center}		\includegraphics[scale=0.35]{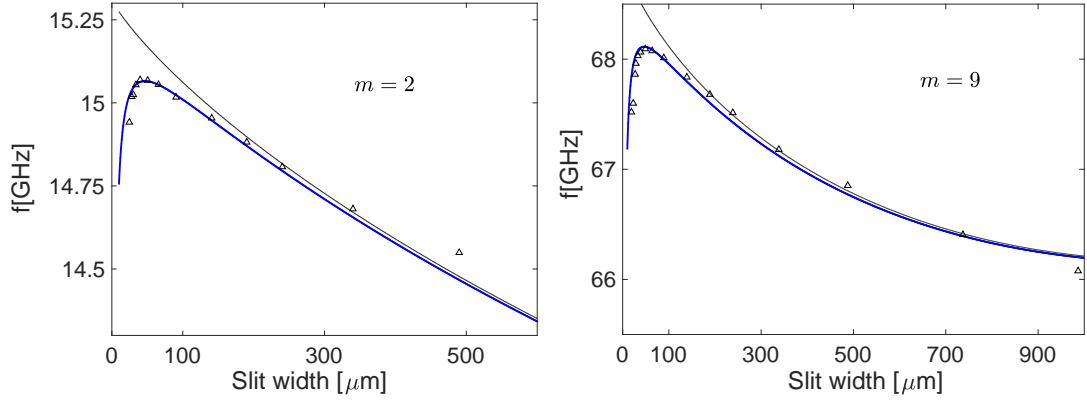}
\caption{Resonance frequencies as a function of slit width for electromagnetic-wave transmission through a single slit in an aluminium plate of thickness $l=19.58\textrm{mm}$. Thick blue lines:  approximation \eqref{single fm}  with $\mathcal{C}$ given by \eqref{C electromagnetic}. Dashed black lines: approximation \eqref{single fm} for the lossless case $\mathcal{C}=0$ \cite{Holley:19}. Symbols: microwave experiments \cite{Suckling:04}. The relative permittivity of the metal is $\epsilon\approx i4.2\times10^7$ and $\epsilon \approx i4.5\times10^6$ for modes $m=2$ and $m=9$, respectively.}
		\label{fig:emex29}
	\end{center}
\end{figure}

\begin{figure}[t!]
	\begin{center}
\includegraphics[scale=0.4]{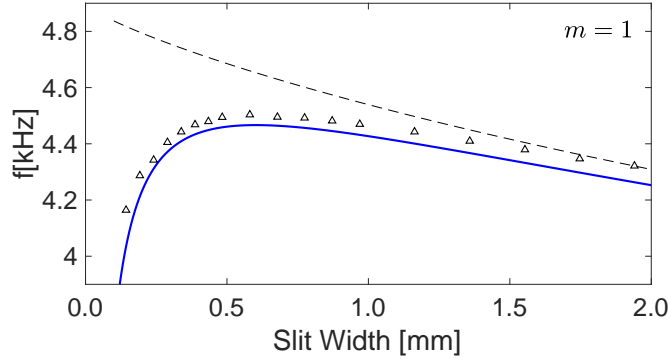}
		\caption{Same as Fig.~\ref{fig:emex29} but for the fundamental resonance of acoustic-wave transmission through a slit in an aluminium plate of thickness $35\mathrm{mm}$. The thick blue line is calculated from \eqref{single fm} with $\mathcal{C}$ calculated from \eqref{C acoustic} for air at $20\,\degree \mathrm{C}$. Symbols: acoustic experiments \cite{Ward:15}.}
		\label{fig:acex1}
	\end{center}
\end{figure}

A closed-form approximation for the resonance frequencies is readily extracted from \eqref{A single}. For the sake of comparison with experimental data, we write this in the form
\begin{equation}\label{single fm}
\frac{\Delta f_m}{f_m}\approx\frac{4h}{\pi}\left(\ln h + \ln(2m) +\gamma_E-1\right)-\frac{h}{2\bar{\Omega}^2}\operatorname{Re}\mathcal{C},
\end{equation}
where $\Delta f_m$ is the dimensional frequency deviation of the $m$th resonant peak from the respective standing-wave frequency $f_m=mc/2l$. In Figs.~\ref{fig:emex29}, and \ref{fig:acex1} we test \eqref{single fm} against GHz-microwave experiments \cite{Suckling:04} and kHz-acoustic experiments \cite{Ward:15}, respectively. The agreement is very good in both cases. In contrast, the lossless approximation with $\mathcal{C}=0$ is not accurate for small slit widths (below approximately $200\mu \textrm{m}$ in the microwave experiments and $1.5\textrm{mm}$ in the acoustic experiments). 

\subsection{Electromagnetic skin layer}\label{ssec:skin}
It remains to derive the effective boundary conditions \eqref{blc} assumed in the slit-region analysis. We begin with the simpler electromagnetic scenario, where we need to consider the internal field $\bar{H}$ in the electromagnetic skin layers, namely at distances $O(\delta)$ from the slit boundaries. Because of the symmetry of the problem, it suffices to consider the skin layer adjacent to the lower boundary of the slit $x=-h$. Since $\delta=O(h^2)$, we define the strained transverse coordinate $X'=(x+h)/h^2$ and consider the domain $X'<0$ ($|y|<1/2$).

Recall the slit-field expansion  \eqref{slit expansion} for $H=\Phi$. In particular, the $O(h^{-1})$ enhancement of the slit field suggests the skin-layer expansion
\begin{equation}
\bar{H}\sim h^{-1}\bar{H}_{-1}(X',y)  \quad \text{as} \quad h\to0.
\end{equation}
The equation governing $\bar{H}_{-1}$ is obtained from the leading-order balance of (\ref{H eqs}b). This gives
\begin{equation}\label{eq skin}
\frac{\partial^2 \bar{H}_{-1}}{\partial X'^2}+\alpha^2\bar{\Omega}^2\bar{H}_{-1}=0 \quad \text{for} \quad X'<0,
\end{equation}
where the complex-valued parameter $\alpha$ was defined in \eqref{alpha def}. From the $O(1/h)$ balance of the continuity condition (\ref{H bcs}a), we find the boundary condition 
\begin{equation}\label{bc skin}
\bar{H}_{-1}=\Phi_{-1} \quad \text{at} \quad X'=0.
\end{equation}
Solving \eqref{eq skin} subject to \eqref{bc skin} and the condition that $\bar{H}_{-1}$ 
decays as $X'\to-\infty$ gives
\begin{equation}\label{barH sol}
\bar{H}_{-1}=e^{-i\alpha\bar{\Omega}X'}\Phi_{-1}.
\end{equation}

Consider now the $O(h)$ balance of the transmission condition (\ref{H bcs}b),
\begin{equation}
\left.\frac{\partial \Phi_{2}}{\partial X}\right|_{X=-1}=\frac{1}{\alpha^2}\left.\frac{\partial \bar{H}_{-1}}{\partial X'}\right|_{X'=0}. 
\end{equation}
Substituting \eqref{barH sol}, we obtain the effective boundary condition
\begin{equation}\label{condM}
\frac{\partial \Phi_{2}}{\partial X}=-\frac{i\bar{\Omega}}{\alpha}\Phi_{-1} \quad \text{at} \quad X=-1.
\end{equation}
The corresponding condition at $X=1$ follows from symmetry.
Using definition \eqref{alpha def} for $\alpha$, \eqref{condM} implies the effective boundary condition \eqref{blc} with $\mathcal{C}$ given by \eqref{C electromagnetic}.

\subsection{Thermoviscous  boundary layer}\label{ssec:thermoviscous}
We next consider the acoustic scenario, where we need to match the thermoviscous boundary layers with the bulk slit region. As a preliminary step, it is convenient to expand the description of the bulk slit region from the single field $p=\Phi$ to include the normalised flow and temperature fields, $\bu=u\be_x+v\be_y$ and $T$, respectively. To this end, we rewrite the slit expansion \eqref{slit expansion} in the form
\begin{equation}\label{P exp}
p= h^{-1}\Phi_{-1}(y) +\Phi_0(y)+h\Phi_{1}(y)+h^2\Phi_2(X,y)+ \cdots \quad \text{as} \quad h\to0.
\end{equation}
Note that the independence of $\Phi_{-1}$, $\Phi_0$ and $\Phi_1$ upon $X$ was determined in \S\ref{ssec:slit} based on the scaling \eqref{slit BC scaling}, which is consistent with the boundary-layer analysis here. Given \eqref{P exp}, the momentum equation \eqref{mom} implies the expansions
\refstepcounter{equation}
$$
\label{uv exp}
u \sim -ih\pd{\Phi_2}{X}, \quad v\sim -i h^{-1}\frac{d\Phi_{-1}}{dy} \quad \text{as} \quad h\to0. 
\eqno{(\theequation{\mathit{a},\mathit{b}})}
$$
Similarly, the energy equation \eqref{energy} implies the temperature expansion
\begin{equation}\label{T exp}
T\sim h^{-1}\Phi_{-1} \quad \text{as} \quad h\to0.
\end{equation}

Consider now the $O(\delta)$ boundary layer adjacent to the lower slit boundary $x=-h$. Since $\delta=O(h^2)$, we 
define the transverse coordinate $X'=(x+h)/h^2$, the boundary-layer domain being $X'>0$ ($|y|<1/2$). The slit-region expansions \eqref{P exp}--\eqref{T exp} suggest expanding the acoustic fields in the boundary layer as
\refstepcounter{equation}
$$
p\sim h^{-1}\bar{P}_{-1}(X',y), \,\, T\sim h^{-1}\bar{T}_{-1}(X',y), \,\, u\sim h\bar{U}_1(X',y), \,\, v\sim h^{-1}\bar{V}_{-1}(X',y).
\eqno{(\theequation{\mathit{a}\!\!-\!\!\mathit{d}})}
$$
The leading $O(h^{-3})$ balance of the $x$ component of the momentum equation \eqref{mom} gives
\begin{equation}
\pd{\bar{P}_{-1}}{X'}=0,
\end{equation}
showing that $\bar{P}_{-1}$ is independent of $X'$. With that,  straightforward matching of the boundary layer and slit-region pressure fields yields 
\begin{equation}\label{P lo}
\bar{P}_{-1}=\Phi_{-1}(y).
\end{equation}

With \eqref{P lo}, the $O(h^{-1})$ balance of the $y$ component of the momentum equation \eqref{mom} can be written  
\begin{equation}\label{V eq}
{\delta'}^2\pd{^2\bar{V}_{-1}}{X'^2}+i\bar{V}_{-1}=\frac{d\Phi_{-1}}{dy},
\end{equation}
where we remind that $\delta'=\delta/h^2$. The no-slip condition \eqref{no slip} gives the boundary condition
\begin{equation}\label{V no slip}
\bar{V}_{-1}=0 \quad \text{at} \quad X'=0.
\end{equation}
Furthermore,  matching  the $v$ velocity component yields the far-field condition
\begin{equation}\label{V far}
\bar{V}_{-1}\to -i\frac{d\Phi_{-1}}{dy} \quad \text{as} \quad X'\to\infty. 
\end{equation}
Solving \eqref{V eq} subject to \eqref{V no slip} and \eqref{V far}, we find
\begin{equation}
\bar{V}_{-1}=i\left(e^{-\frac{1-i}{\sqrt{2}}\frac{X'}{\delta'}}-1\right)\frac{d\Phi_{-1}}{dy}.
\end{equation}

The leading $O(h^{-1})$ balance of the energy equation \eqref{energy}, with \eqref{P lo}, reads 
\begin{equation}\label{T eq}
\mathrm{Pr}^{-1}{\delta'}^2\pd{^2\bar{T}_{-1}}{X'^2}+i\bar{T}_{-1}=i\Phi_{-1}.
\end{equation}
From \eqref{no temp}, we find the boundary condition
\begin{equation}\label{T bc}
\bar{T}_{-1}=0.
\end{equation}
Furthermore, leading-order matching of the temperature field gives the far-field condition
\begin{equation}\label{T far}
\bar{T}_{-1}\to \Phi_{-1} \quad \text{as} \quad X'\to\infty. 
\end{equation}
Solving \eqref{T eq} subject to \eqref{T bc} and \eqref{T far}, we find
\begin{equation}
\bar{T}_{-1}=\left(1-e^{-\frac{1-i}{\sqrt{2}}\frac{X'\sqrt{\mathrm{Pr}}}{\delta'}}\right)\Phi_{-1}.
\end{equation}

Consider now the leading $O(h^{-1})$ balance of the continuity equation \eqref{cont},
\begin{equation}\label{cont lo}
\pd{\bar{U}_1}{X'}=-\pd{\bar{V}_{-1}}{y}+i\bar{\Omega}^2\bar{P}_{-1}+i(\gamma-1)\bar{\Omega}^2\left(\bar{P}_{-1}-\bar{T}_{-1}\right),
\end{equation}
which is supplemented by the boundary condition [cf.~\eqref{no slip}]
\begin{equation}\label{U1 bc}
\bar{U}_1=0 \quad \text{at} \quad X'=0.
\end{equation}
Integrating \eqref{cont lo} using \eqref{U1 bc}, followed by taking the limit ${X'}\to\infty$ using the slit-region Helmholtz equation \eqref{leading slit helm}, yields 
\begin{equation}
\bar{U}_1\to -\delta'\bar{\Omega}^2\frac{1-i}{\sqrt{2}}\left(1+\frac{\gamma-1}{\sqrt{\mathrm{Pr}}} \right)\Phi_{-1}\quad \text{as} \quad X'\to\infty.
\end{equation}
Matching the velocity component $u$ then furnishes the effective boundary condition
\begin{equation}\label{eff condition phi2}
\pd{\Phi_2}{X}=-\delta'\bar{\Omega}^2\frac{1+i}{\sqrt{2}}\left(1+\frac{\gamma-1}{\sqrt{\mathrm{Pr}}}\right) \Phi_{-1} \quad \text{at} \quad X=-1.
\end{equation}
The corresponding condition at $X=1$ follows from symmetry. Together these give expression \eqref{C acoustic} for the parameter $\mathcal{C}$. 

We emphasise that the effective condition \eqref{eff condition phi2} relies on the one-dimensional nature of the wave propagation in the bulk slit region and the related independence of $\Phi_{-1},\Phi_0$ and $\Phi_1$ on the transverse coordinate $X$. When the bulk field is multi-dimensional, the viscous effect gives a contribution to the effective boundary condition proportional to the surface Laplacian of the bulk pressure, rather than the bulk pressure itself \cite{Brandao:20O}. Given   \eqref{leading slit helm}, in the present case $\Phi_{-1}$ and its surface Laplacian are in fact proportional.  

\begin{figure}[t!]
	\begin{center}		\includegraphics[scale=0.38]{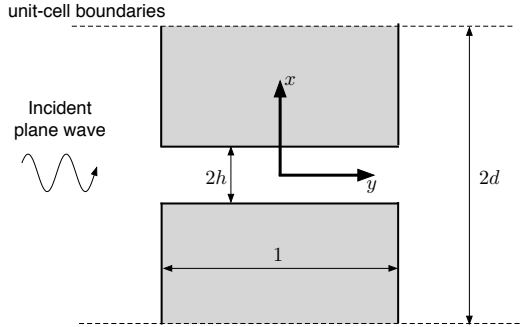}		\caption{Dimensionless schematic of the problem of transmission through a periodic array of slits.}
		\label{fig:array}
	\end{center}
\end{figure}
\section{Extension to a periodic slit array}\label{sec:array}
\subsection{Formulation}
Consider now a more complex configuration where the plate is decorated with an infinite array of slits, say of dimensional period  $2dl$. Fig.~\ref{fig:array} shows the corresponding dimensionless geometry, with the origin of the Cartesian coordinates ($x$,$y$) placed at the centre of an arbitrary slit. For a normally incident plane wave, the electromagnetic and acoustic fields possess the same periodicity as the array. Accordingly, the problem can be restricted to the unit-cell $|x|<d$, say, with periodicity conditions applied on the cell boundaries $x=\pm d$.  

\subsection{Analysis and frequency response}
We shall extensively build on the near-resonance theory developed in \S\ref{sec:single} for a single slit. Thus, in the exterior regions, the periodicity of the problem suggests that the leading $O(1)$ fields $\varphi^{\pm}$ generalise as [cf.~\eqref{exterior left} and \eqref{exterior right}]
\refstepcounter{equation}
$$
\label{plor} 
\varphi^{-}= e^{i\bar{\Omega}y}+e^{-i\bar{\Omega}(1+y)}+Q^{-}\sum_{n=-\infty}^{\infty}\mathcal{H}_0\left(\bar{\Omega}r_{n}^{-}\right), \quad \varphi^{+}= Q^{+}\sum_{n=-\infty}^{\infty}\mathcal{H}_0\left(\bar{\Omega}r_{n}^{+}\right),
\eqno{(\theequation{\mathit{a},\mathit{b}})}
$$
where we define the shifted radial coordinates $r_{n}^{\pm}=\sqrt{(x-2nd)^2+(y\mp1/2)^2}$. We next couple these exterior fields to the bulk slit  field, using the matching formulae \eqref{Q in A} and \eqref{relation 2}. For this purpose, we consider just the slit in the unit-cell $|x|<d$. The expansion of the wave field in the slit region has exactly the same form as in the single-slit case, with $\mathcal{A}$ denoting the complex amplitude of the resonantly excited standing wave at $O(1/h)$. Thus, analogously to \eqref{Phi0 at ends}, we find the end values of the slit field at $O(1)$ in the form
\begin{equation}\label{Phi0 ends array}
\Phi_0(\pm 1/2)=iQ^{\pm}\beta+Q^{\pm}\left[1+\frac{2i}{\pi}\left(\ln\frac{\bar{\Omega}h}{2}+\gamma_E\right)+\sigma(\bar{\Omega}d)\right]+(1\mp1)e^{-\frac{i\bar{\Omega}}{2}},
\end{equation}
in which the function $\sigma(\bar\Omega d)$ is formally given by the conditionally convergent lattice sum
\begin{equation}\label{sigma def}
\sigma(\bar\Omega d)=\sum_{n\neq 0}\mathcal{H}_0\left(2\bar\Omega |n|d\right).
\end{equation}
For computation, this sum is transformed into the absolutely converging sum (see \cite{Linton:98})
\begin{equation}\label{sigma abs}
\sigma(\bar\Omega d)=-1-\frac{2i}{\pi}\left(\gamma_E+\ln\frac{{\bar\Omega}d}{2\pi}\right)-\frac{i}{\chi_0}-i\sum_{n\neq0}\left(\frac{1}{\chi_n}-\frac{1}{\pi |n|}\right),
\end{equation}
wherein
\begin{gather}
\chi_n= \Bigg\{\begin{array}{c}\sqrt{\pi^2n^2-\bar{\Omega}^2d^2},\quad |n|\geqslant \bar{\Omega}d/\pi, \\ -i\sqrt{\bar{\Omega}^2d^2-\pi^2n^2},\quad |n|<\bar{\Omega}d/\pi.\end{array}
\end{gather}
Along the lines of the single-slit analysis, we solve \eqref{A loss}, \eqref{Q in A} and \eqref{Phi0 ends array} for the complex amplitude $\mathcal{A}$. We thereby find the generalised near-resonance approximation
\begin{equation}\label{A array}
\mathcal{A}=\frac{1/\bar{\Omega}}{\Omega'+\frac{\mathcal{C}}{4\bar{\Omega}^2}-\frac{2}{\pi}\left(\ln\frac{2\bar{\Omega} h}{\pi}+\gamma_E-1\right)+i(1+\sigma(\bar{\Omega}d))}\times \left\{\begin{array}{c}i \\ -1 \end{array}\right\}.
\end{equation}
The corresponding formulae for $Q^{\pm}$ follow from \eqref{Q in A}.

Comparing \eqref{A array} with \eqref{A single}, we see that the generalisation to a periodic array enters through the complex-valued function $\sigma(\bar{\Omega} d)$. Its real part contributes to radiative damping of the slit mode, while its imaginary part shifts the resonances to lower frequencies. In the above analysis, we tacitly held $d$ fixed; namely, we assumed that the periodicity is comparable with the wall thickness and the wavelength. It can be verified, however, that the theory holds as a leading-order near-resonance approximation for $\delta\ll h\ll d$. For large $d$, $\sigma$ attenuates and \eqref{A array} reduces to \eqref{A single}. 
For small $d$ (subwavelength periodicity), $\sigma\sim (\bar{\Omega}d)^{-1}$, thus $\sigma$ is real and large implying that the slit fields are more strongly damped by radiation than in the single-slit scenario. Since $\mathcal{C}=O(\delta/h^2)$, the distinguished limit \eqref{delta p def} becomes $\delta=O(h^2/d)$; thicker boundary layers are required for dissipative and radiative losses to be comparable.  
 \begin{figure}[t!]
	\begin{center}
\includegraphics[scale=0.5]{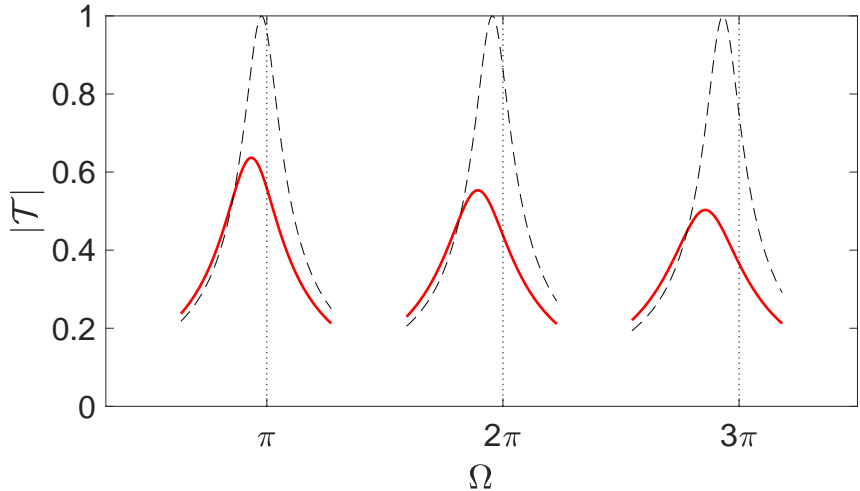}
		\caption{Modulus of the transmission coefficient for acoustic transmission through a plate of thickness $l=19.8\textrm{mm}$ decorated by a periodic array of slits of width $w=2hl$ and period $2dl= w+2.91\textrm{mm}$ (dimensions from \cite{Ward:15}), for $h=0.01$. Solid line: approximation \eqref{RT} with $\mathcal{C}$ given by \eqref{C acoustic} for air at $20\,\degree\mathrm{C}$. Dashed line: approximation \eqref{RT} for the lossless case $\mathcal{C}=0$.}
		\label{fig:tc}
	\end{center}
\end{figure}

Consider next the field at large distances from the plate. To this end, we shall use the asymptotic relation (see \cite{Linton:98}) 
\begin{equation}\label{linton far}
\sum_{n=-\infty}^{\infty}\mathcal{H}_0\left(\bar{\Omega} r_{n}^{\pm}\right)\sim\sum_{n=-\lfloor \bar{\Omega}d/\pi\rfloor}^{\lfloor \bar{\Omega}d/\pi\rfloor}\frac{e^{i\left\{\frac{\pi nx}{\bar{\Omega} d}\pm \sqrt{\bar{\Omega}^2d^2-\pi^2n^2}\frac{y\mp1/2}{d}\right\}}}{\sqrt{\bar{\Omega}^2d^2-\pi^2 n^2}}\quad\text{as}\quad y\to\pm\infty.
\end{equation}
In particular, when the period of the array is smaller than the wavelength ($\bar{\Omega}d<\pi$), the finite sum on the right-hand side of \eqref{linton far} reduces to a single plane-wave term. In that case, we find from \eqref{plor} the far-field behaviours 
\refstepcounter{equation}
$$
\varphi^{-}\sim e^{i\bar{\Omega}y}+\mathcal{R}e^{-i\bar{\Omega}y}\quad\text{as}\quad y\to-\infty, \quad \varphi^{+}\sim \mathcal{T}e^{i\bar{\Omega}y}\quad\text{as}\quad y\to\infty,
\eqno{(\theequation{\mathit{a},\mathit{b}})}
$$
where we define the complex-valued reflection and transmission coefficients
\refstepcounter{equation}
$$
\label{RT}
\mathcal{R}=e^{-i\bar{\Omega}}+\frac{Q^{-}}{\bar{\Omega}d}e^{-\frac{i\bar{\Omega}}{2}}, \quad \mathcal{T}=\frac{Q^{+}}{\bar{\Omega}d}e^{-\frac{i\bar{\Omega}}{2}}.
\eqno{(\theequation{\mathit{a},\mathit{b}})}
$$
\begin{figure}[b!]
	\begin{center}		\includegraphics[scale=0.45]{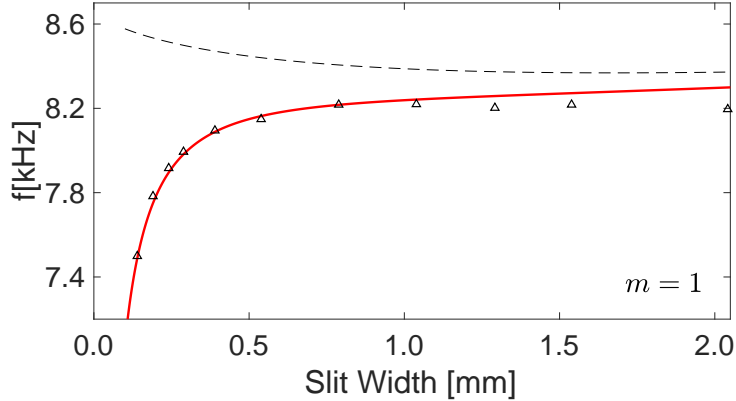}
		\caption{Frequency of fundamental resonance as a function of slit width $w=2hl$ for acoustic-wave transmission through a periodic array of slits, of period $2dl=w+2.91\textrm{mm}$, in an aluminium plate of thickness $l=19.8\textrm{mm}$. Thick red line:  approximation \eqref{array fm} with $\mathcal{C}$ given by \eqref{C acoustic} for air at $20\,\degree\mathrm{C}$. Dashed black line: approximation \eqref{array fm} for the lossless case $\mathcal{C}=0$. Symbols: acoustic experiments \cite{Ward:15}.}
		\label{fig:pacex1}
	\end{center}
\end{figure}

In Fig.~\ref{fig:tc} we plot $|\mathcal{T}|$ for $h=0.01$ and with the remaining geometric dimensions from the slit-array acoustic experiments in \cite{Ward:15}. In particular, we compare the lossless case $\mathcal{C}=0$ and the thermoviscous-acoustic scenario where $\mathcal{C}$ is provided by \eqref{C acoustic} for air at $20\,\degree\mathrm{C}$. In the lossless case, we observe perfect transmission at frequencies  shifted slightly downwards from the unperturbed standing-wave frequencies. The thermal and viscous effects are seen to diminish the transmission peaks and shift these to yet lower frequencies. 

\subsection{Comparison with slit-array acoustic experiments}\label{ssec:exp2}
A closed-form approximation for the resonance frequencies can be readily extracted from \eqref{A array}. We write this as
\begin{equation}\label{array fm}
\frac{\Delta f_m}{f_m}\approx\frac{4h}{\pi}\left(\ln h + \ln 2m +\gamma_E-1\right)-\frac{h}{2\pi^2m^2}\operatorname{Re}\mathcal{C}+2h\operatorname{Im}\sigma(md),
\end{equation}
where $\Delta f_m$ is again the deviation of the $m$th resonance frequency from the corresponding unperturbed standing-wave frequency $f_m$.
In Fig.~\ref{fig:pacex1} we test \eqref{array fm} against the acoustic slit-array experiments in \cite{Ward:15}. The agreement is very good. 

\section{Concluding remarks}\label{sec:conc}
We have developed an asymptotic theory describing resonant electromagnetic- and acoustic-wave transmission through slitted plates. The theory provides simple scaling rules and analytical approximations that accurately capture both diffractive and dissipative effects, as demonstrated through a comparison with experimental results in the literature. We hope that this work showcases the power of scaling arguments, matched asymptotics and near-resonance expansions for modelling structured-wave devices. Thus, the present theory can be readily adapted to other configurations where resonant slits are used to manipulate or absorb waves \cite{Moleron:16,Ward:16,Schnitzer:17}. A similar approach could also be developed for three-dimensional hole resonators \cite{Ward:19}.  An analogous theory for acoustic Helmholtz resonators embedded in a wall and arrays thereof has recently been developed by two of us \cite{Brandao:20H}.

We conclude with comments regarding the generalised electromagnetic-acoustic analogy identified in this paper. First, this analogy is asymptotic, in contrast to the exact analogy between the respective lossless problems. Second, the generalised analogy relies on the findings that (i) boundary-layer effects are only appreciable in the slit region and (ii) the propagation within the bulk of the slit is effectively one-dimensional. This suggests that the analogy could be extended to other slit configurations where boundary-layer effects are important. In general, however, thermoviscous boundary layers and electromagnetic skin layers are described by different types of effective boundary conditions. 

\bibliographystyle{apsrev4-1}
\bibliography{refs}

\end{document}